\documentclass[submit]{smj}
\usepackage{times,graphicx,amsmath,bm,algorithm,algorithmic,setspace}

\Author{Xin-Jian Xu\Affil{1}, Cheng Chen\Affil{1}, and J. F. F. Mendes\Affil{2}}
\AuthorRunning{Xin-Jian Xu \textrm{et al.}}

\Affiliations{

\item Department of Mathematics,
      Shanghai University,
      Shanghai,
      China
\item Department of Physics \& I3N,
      University of Aveiro,
      Aveiro,
      Portugal
}   %% end \Affiliations

\CorrAddress{Xin-Jian Xu,
             Department of Mathematics,
             Shanghai University,
             Shanghai 200444,
             China}
\CorrEmail{xinjxu@shu.edu.cn}
\CorrPhone{(+86)\;21\;6613\;4715}
\CorrFax{(+86)\;21\;6613\;4715}

\Title{Efficient hypothesis testing for community detection in heterogeneous networks}
\TitleRunning{Hypothesis testing for community detection}

\Abstract{
Identifying communities in networks is a fundamental and challenging problem of practical importance in many fields of science. Current methods either ignore nodal heterogeneity or assume prior knowledge of the number of communities. Here we propose an efficient hypothesis test for community detection based on quantifying dissimilarities between graphs. Given a random graph, the null hypothesis is that it is of degree-corrected Erd\"{o}s-R\'{e}nyi type. We compare the dissimilarity between them by a measure incorporating the vertex distance distribution, the clustering coefficient distribution and the alpha-centrality distribution, which is used for our hypothesis test. We also design a two-stage bipartitioning algorithm to uncover the number of communities and the corresponding structure simultaneously. Experiments on synthetic and real networks show that our method outperforms state-of-the-art ones.
}


\begin{document}

%%% Title page
%%% -------------------------------------------------------------------------------
%%% Use command\maketitle to produce the title page.
\maketitle

%%% Main text
%%% ------------------------------------------
\section{Introduction}

Over the two decades, complex networks have been used to describe a variety of complex systems. With vertices representing systematic units and edges connected them representing interactions, we obtain real networks spanning many different fields \citep{Barabasi15book}. One of the pertinent characteristics of these real networks is that they display community structure, i.e., vertices are organized into groups. The task of group identification is known as community detection \citep{Fortunato10pp}, in many ways similar to graph clustering. Although a large number of algorithms for community detection have been proposed, such as clustering algorithms \citep{Maqbool04proc, Newman06pnas}, modularity-based algorithms \citep{Clauset04pre, Blondel08jsm}, dynamic algorithms \citep{Palla05nat, Rosvall07pnas}, etc, a single community detection algorithm usually fails to perform in all kind of networks \citep{Hric14pre}, and therefore a general efficient method remains demanding.

From the probability perspective, vertices in the same community have higher possibility to be connected than those in the different communities. Thus, the stochastic block model (SBM) \citep{Holland83sn, Abbe18jmlr} has been adopted to detect communities, which provides a theoretical framework for the study of the detection threshold and corresponding algorithms. A seminal paper by \citet{Decelle11pre} conjectured the phase transition for community detection at the Kesten-Stigum threshold, which triggered several studies on different transition thresholds for different recovery conditions \citep{Abbe15proc, Mossel15ptrf, Abbe16ieee}. On the other hand, many algorithms for the SBM have been proposed depending on the specific research question or the particular system at study, such as spectral methods \citep{Rohe11as, Angel15tams, Jin15as, Sarkar15as}, semi-definite programming methods \citep{Abbe16ieee, Guedon16ptrf}, profile-likelihood maximization \citep{Bickel09pnas} and pseudo-likelihood maximization \citep{Amini13as}. In particular, \citet{Peixoto12prl, Peixoto17pre} considered the number of edges instead of their connecting probabilities and provided a microcanonical view on the SBM.

The standard SBM assumes that vertices in the same community are stochastically equivalent and have the same expected degree, which violates real networks due to the present of hubs with many connections than other vertices in their community \citep{Artico20prsa}. In view of this, \citet{Karrer11pre} proposed the degree-corrected SBM (DCSBM), multiplying the probability of an edge between vertices $i$ and $j$ by the product of vertex-specific \lq\lq degree parameters\rq\rq. Following this idea, several studies have devoted to the DCSBM in community detection. \citet{Zhao12as} established the general theory for checking consistency of community detection under the DCSBM and compared different community detection criteria under both the SBM and DCSBM. \citet{Chen18as} proposed a method based on a convex programming relaxation of modularity maximization and designed a weighted $\ell_1$-norm $k$-medoids algorithm. In contrast, \citet{Gao18as} derived the misclassification proportion by calculating asymptotic minimax risks, which depends on the degree parameter, the community size, and the connecting parameter. All of these algorithms assume that the number of communities is prior known. %in contrast to theoretical assumption that this number grows with network size~\cite{hu20jasa}.

In practice, all information we can get is the vertex set and the edge set, i.e., vertices of which are connected to each other and which are not. Thus, determining the number of communities is a challenging issue. To the best of our knowledge, existing approaches only focused on the SBM. One direction is detecting the optimal community structure from different numbers of communities first, and then penalizing the model parameters with the minimum description length \citep{Rosvall07pnas}, the Akaike information criterion \citep{Burnham04book}, or the Bayesian information criterion \citep{Latouche12sm}. Another direction is developing hypothesis tests for determining the number of communities, from the perspectives of asymptotic consistency \citep{Zhao11pnas} or the principal eigenvalue of a normalized adjacency matrix \citep{Bickel16jrss}. However, the methods of the both directions either needs much time for large networks or may underestimate or overestimate the number of communities.

The goal of this paper is to simultaneously uncover the number of communities and the corresponding structure in heterogeneous networks in an efficient way. To this end, we propose a novel hypothesis test based on graph dissimilarity, which is a function of the vertex distance distribution, the clustering coefficient distribution, and the alpha-centrality distribution. The null hypothesis is assuming that the original network is a one-block DCSBM, i.e., the degree-corrected Erd\"{o}s-R\'{e}nyi graph (DCERG), from which one can estimate the connecting parameter and the degree parameter. Then we compute the dissimilarity between the original network and the posterior DCERG, and use the kernel density estimation (KDE) to formulate the dissimilarity distribution among DCERGs generated by same parameters. If the hypothesis is rejected, we split the network by the bipartitioning algorithm until each sub-graph accepts the hypothesis. Our method circumvents the calculation of the eigenvalues of the adjacency matrix, hence much efficacy.

\section{The hypothesis test}

As a probabilistic generative model for random graphs with community structure, the SBM combines the strict block model with a stochastic element and serves as a benchmark for the task of recovering community structure from network data. Let $G$ be a simple graph with $N$ vertices grouped into $K$ blocks, then each vertex $i \in [N]$ is a member of exactly one block determined by the prior probability $p_j$ with $j \in [K]$, which satisfies the normalization $\sum_{j=1}^{K}p_j=1$. The basic idea of the SBM is that the neighborhood relations of each vertex depend only on the probabilities given by the block memberships. Let $w_{st}$ be the connecting probability between one vertex in block $s$ and the other vertex in block $t$, then $\bm{W}$ is a $K \times K$ matrix. Now we can write the conditional expectation of the adjacency matrix $\bm{A}$ given the block assignments $b$:
\begin{equation}\label{sbm}
 \text{E}(a_{ij}|b)=w_{b_i,b_j},
\end{equation}
where $a_{ij}=1$ if there is an edge from $i$ to $j$, and $0$ otherwise. When all the $b_i$ are identical, the SBM reduces to the classic Erd\"{o}s-R\'{e}nyi graph (ERG), and no meaningful reconstruction of the communities is possible. Given an instance of real networks, one can fit the model by maximizing this expectation with respect to node labels $b_i$, and the goal of the community detection problem is to recover these labels.

From the definition of the standard SBM, two vertices assigned into the same block have the same probabilities to connect other vertices. As a result, the SBM does not allow for the existence of \lq\lq hubs\rq\rq, and the maximum of the log likelihood function based on it would split the graph into a group composed of high-degree vertices and another of low-degree vertices. To solve this problem, \citet{Karrer11pre} proposed the DCSBM, which replaces Eq.~\eqref{sbm} with
\begin{equation}\label{dcsbm}
 \text{E}(a_{ij}|\theta,b)=\theta_i\theta_jw_{b_i,b_j},
\end{equation}
where $\theta_i$ is a degree parameter associated with vertex $i$ reflecting its individual propensity to form ties. As $\theta_i$ control the expected degree of vertex $i$, it has to satisfy a constraint to be identifiable, which was set to $\sum \theta_i \delta_{b_i,s}=1$ for all blocks $s$. The DCSBM allows heterogeneity inside each block and the likelihood to observe at least one edge between vertices is in the degree-corrected case, so $\theta_i$ equals to the probability that an edge connected to the block to which $i$ belongs lands on $i$ itself.

A challenge for both the SBM and the DCSBM is the requirement of the priori knowledge about the actual number of blocks. The hypothesis test is a promising approach to overcome this limitation. In essence, determining whether a DCSBM have $K$ or $K+1$ blocks can be thought of as inductively deciding whether there is one block or two. This inspires us with null hypothesis: the network is a one block DCSBM, i.e., the DCERG. The expection of the adjacency matrix of the DCERG is given by
\begin{equation}\label{dcsbm}
 \text{E}(\bm{A})=\bm{D}\bm{Z}\bm{D}
\end{equation}
with $\bm{D}=\text{diag}(\theta_1,\theta_2,\cdots,\theta_N)$ and $\bm{Z}=Nw\bm{e}\bm{e}^T-w\bm{I}$, where $\bm{e}$ is a vector with $e_i=1/\sqrt{N}$ for $i \in [N]$ and $\bm{I}$ is the identity matrix. Assuming that the graph is generated by the DCERG, we need to estimate $\theta$ and $w$. The former is given by
\begin{equation}\label{dcsbm}
 \hat{\theta}_i=\frac{k_i}{\sum_{i=1}^N k_i},
\end{equation}
where $k_i=\sum_{j=1}^{N} a_{ij}$ is the degree of vertex $i$. While the later can be obtained as
\begin{equation}\label{dcsbm}
 \hat{w}=\frac{\sum_{i=1,j=1}^N e_{ij}}{N(N-1)}
\end{equation}
with $e_{ij}=\theta_i^{-1}a_{ij}\theta_j^{-1}$.

Now the problem becomes to distinguishing the DCSBM($N,p,W,\theta$) and the DCERG($N,\hat{w},\hat{\theta}$). In general, measuring the structural dissimilarity of large graphs is a challenging undertaking because of the often unfavorable computational complexity of the analysis methods \citep{Schieber17nc}. Although the literature on this topic is abundant, existing studies have focused so far on networks with simple structure while degree heterogeneity and community structure were usually ignored~\citep{Emmert16is}. To overcome this limitation, we have proposed a precise and efficient measure to quantify dissimilarities between graphs from the perspective of probability distribution functions \citep{Xu22pa}:
%\begin{eqnarray}\label{dissim}
%D(G,G') &=& \gamma_1\sqrt{\frac{\mathcal{J}(Q_l(G),Q_l(G'))}{\log{2}}}\notag\\
 %  &+& \gamma_2\sqrt{\frac{\mathcal{J}(Q_c(G),Q_c(G'))}{\log{2}}}\notag\\
 %  &+& \gamma_3\sqrt{\frac{\mathcal{J}(Q_{\alpha}(G),Q_{\alpha}(G'))}{\log{2}}},
%\end{eqnarray}
\begin{equation}\label{dissim}
 D(G,G') = \gamma_1\sqrt{\frac{\mathcal{J}(Q_l(G),Q_l(G'))}{\log{2}}} +\gamma_2\sqrt{\frac{\mathcal{J}(Q_c(G),Q_c(G'))}{\log{2}}} +\gamma_3\sqrt{\frac{\mathcal{J}(Q_{\alpha}(G),Q_{\alpha}(G'))}{\log{2}}},
\end{equation}
where $\gamma_1$, $\gamma_2$ and $\gamma_3$ are arbitrary weights of the terms satisfying $\gamma_1+\gamma_2+\gamma_3=1$. $Q_l(G)=\{q_l(i)\}=\{\sum_{i=1}^N n_{ik}/N(N-1)\}$ is the vertex distance distribution of the network with $n_{ik}$ being the number of nodes at distance $k$ from $i$. $Q_c(G)=\{q_c(i)\} =\{[\pi_c;N-\sum_{i=1}^N\pi_c(i)]/N\}$ is the vertex clustering coefficient distribution of the network with $\pi_c$ being ordered by increasing values of nodal clustering coefficient. $Q_{\alpha}(G)=\{q_{\alpha}(i)\}=\{[\pi_{\alpha};N-\sum_{i=1}^N\pi_{\alpha}(i)]/N\}$ is the vertex centrality distribution of the network with $\pi_{\alpha}$ being ordered by increasing values of nodal $\alpha$-centrality. $\mathcal{J}(\bm{q}_1,\bm{q}_2)=\frac{1}{2}\sum_i q_1(i)\ln[2q_1(i)/(q_1(i)+q_2(i))]+\frac{1}{2}\sum_i q_2(i)\ln[2q_2(i)/\sum_i (q_1(i)+q_2(i))]$ is the Jensen-Shannon divergence. Defined in this way, $D$ captures both global and local dissimilarities of two graphs. Moreover, it is easy to confirm that $D \in [0,1)$. Finally, to calculate the $P$-value to accept or reject the null hypothesis, we need the distribution of $D$, which can be obtained by the KDE.

On the base of the above discussion, we present the following two-stage hypothesis test algorithm.

\begin{algorithm}[h]
\setstretch{1.3}
\caption{Hypothesis test algorithm}\label{algor}
\begin{algorithmic}[1]
\STATE $\bm{A}$ $\leftarrow$ adjacency matrix of $G$\\
    \quad $\hat{\theta}_i \leftarrow \frac{k_i}{\sum_i^N k_i}$, $\hat{w} \leftarrow \frac{\sum_{i=1,j=1}^N E_{ij}}{N(N-1)}$\\
\quad For $i=1,2,\cdots,50$\\
    \quad \quad $G_i \leftarrow \text{DCERG}(N,\hat{w},\hat{\theta}$), $\overline{D}=\sum D(G,G_i)/50$\\
    \quad For all $i\neq j$\\
    \quad \quad $D_{ij}$ $\leftarrow$ $D(G_i,G_j)$\\
    \quad $\hat{P}(D(G,\text{DCERG}(N,\hat{w},\hat{\theta}))) \leftarrow \text{KDE}(D_{ij})$\\
\STATE pval $\leftarrow$ $\hat{P}(\overline{D}>D)$\\
    \quad If pval $<$ significant level $\alpha$\\
    \quad \quad i) For each edge $e_{ij}$\\
        \quad \quad \quad \quad compute the edge betweenness $B_{ij}$ and edge clustering coefficient $C_{ij}$\\
        \quad \quad \quad \quad $L_{ij} \leftarrow \beta_1B_{ij}-\beta_2C_{ij}$\\
        \quad \quad \quad \quad remove edge $e_{ij}$ with $L=\text{max}(L_{ij})$\\
    \quad \quad ii) If the graph is connected\\
         \quad \quad \quad \quad go back to i)\\
    \quad \quad \quad Else\\
         \quad \quad \quad \quad Output $G_1$, $G_2$\\
    \quad \quad \quad End if\\
    \quad Else\\
    \quad \quad Output $G$\\
    \quad End if
\end{algorithmic}
\end{algorithm}

We remark that the former definition of the edge betweenness, $B_{ij}= \sum l(s,t|e_{ij})/l(s,t)$, does not consider the local assortativity of communities. To correct it, we take edge clustering coefficients into account by defining the edge inter-communities measure $L_{ij}=\beta_1B_{ij}-\beta_2C_{ij}$ with $C_{ij}=\Delta_{ij}/\text{min}(k_i-1,k_j-1)$.

\begin{figure}[t]
 \centering
 \includegraphics[width = 0.45\linewidth]{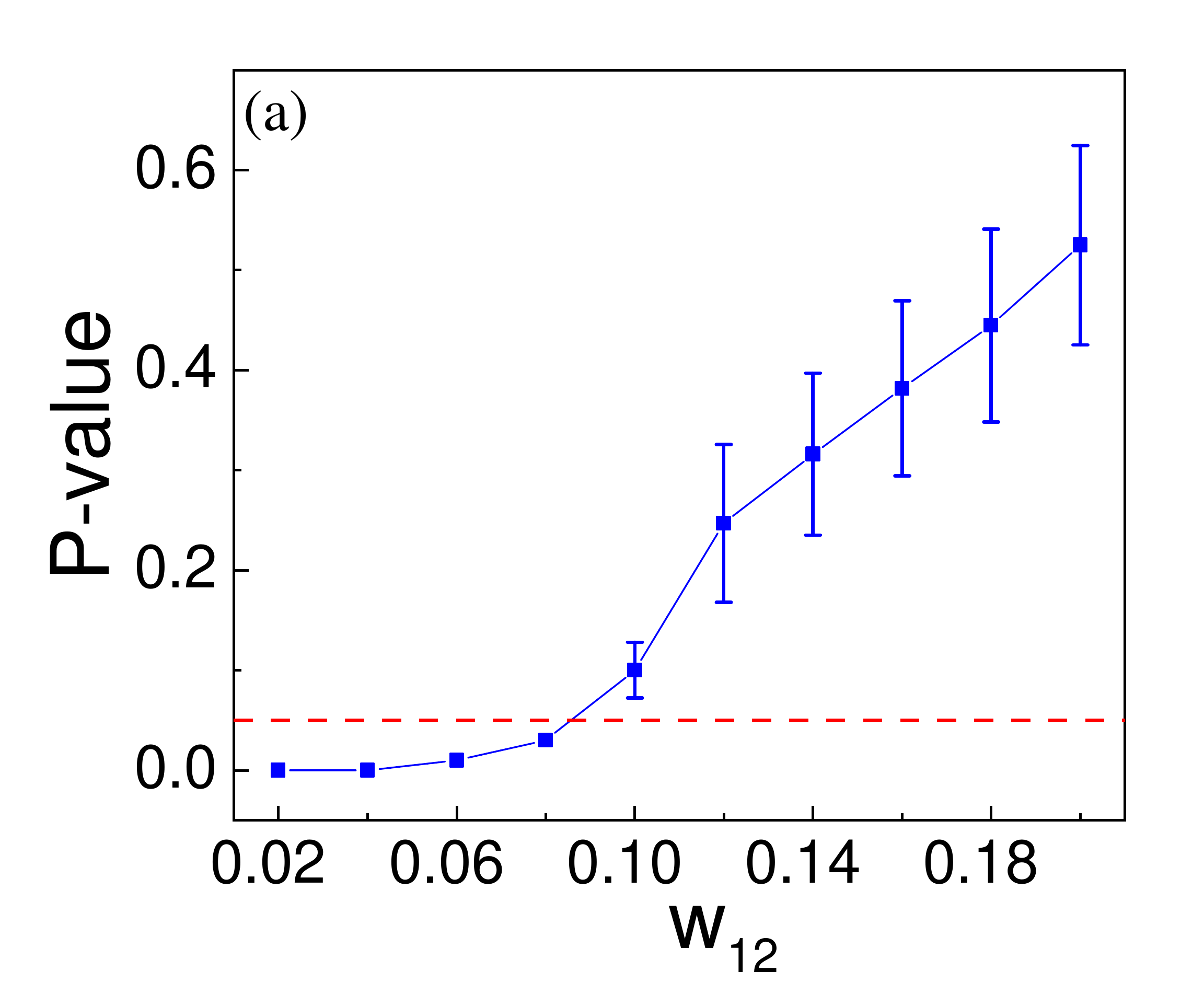}
 \includegraphics[width = 0.45\linewidth]{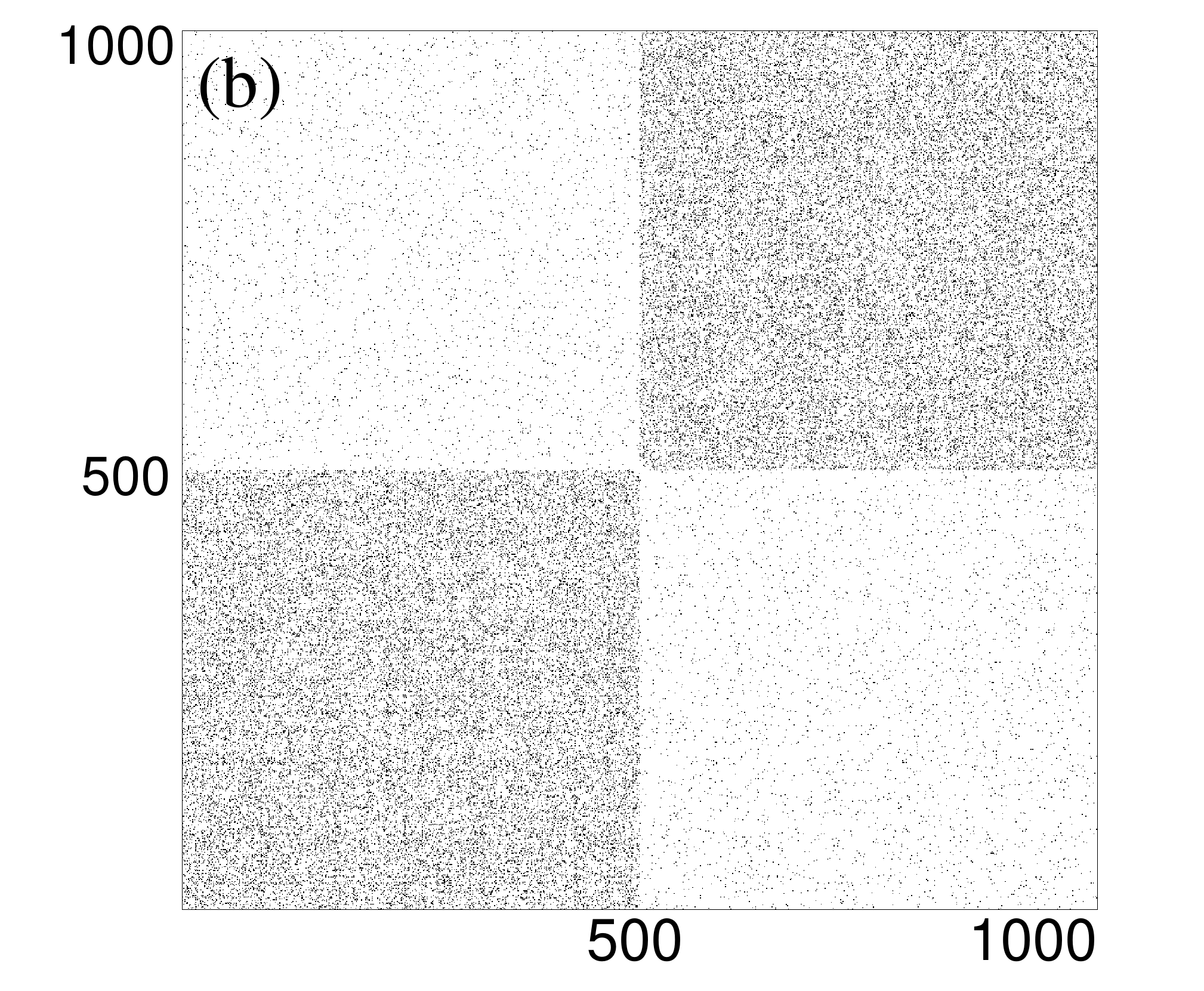}
 \caption{Simulation results of the hypothesis test algorithm for the balanced two-block DCSBM: $P$-value as a function of the connecting parameter $w_{12}$ (a) and the Illustration of the adjacency matrix for $w_{12}=0.02$ (b). The dashed line corresponds to the significant level $\alpha=0.05$}
 \label{figdcsb}
\end{figure}

\section{Application to block models}

To test the performance of our algorithm, we first apply it to the balanced DCSBM, namely, the size of each block are identical. Specially, we set $N=1000$, $K=2$, and $w_{11}=w_{22}=0.2$. The degree parameters $\theta_i$ is drawn from the adjusted normal distribution $\theta \sim \left(|\text{Normal}(0,0.25)| + 1 - \frac{1}{\sqrt{2\pi}} \right)$, following the right-skewed character. Other distributions are also investigated (not shown here). The mean of the distribution is set to $\text{E}(\theta)=1$ without loss of generality. The graph generation is a straightforward implementation of the block model: (i) drawing a Poisson-distributed number of edges for each pair of blocks $1$ and $2$ with $w_{12}=w_{21}$ (or $w_{11}/2=w_{22}/2$ for the same block; and (ii) assigning each end of an edge to a vertex in the proper block with probability $\theta_i$. Because we wish to be able to vary the level of community structure in generated networks, we increase $w_{12} (=w_{21})$ from $0.02$ to $0.2$ in steps of $0.02$. The error bars on $P$-values are computed from $100$ random runs. In essence, a larger $P$-value simply means that the hypothesis test considers the graph to be close to an ERG. As shown in Figure~\ref{figdcsb}(a), the $P$-value increases with $w_{12}$, implying that the network is losing its block structure. To visualize the block structure uncovered by our algorithm, Figure~\ref{figdcsb}(b) illustrates the adjacency matrix for $w_{12}=0.02$, whose rows and columns are ordered. One notices the perfect clustering.

\begin{figure}[t]
 \centering
 \includegraphics[width = 0.45\linewidth]{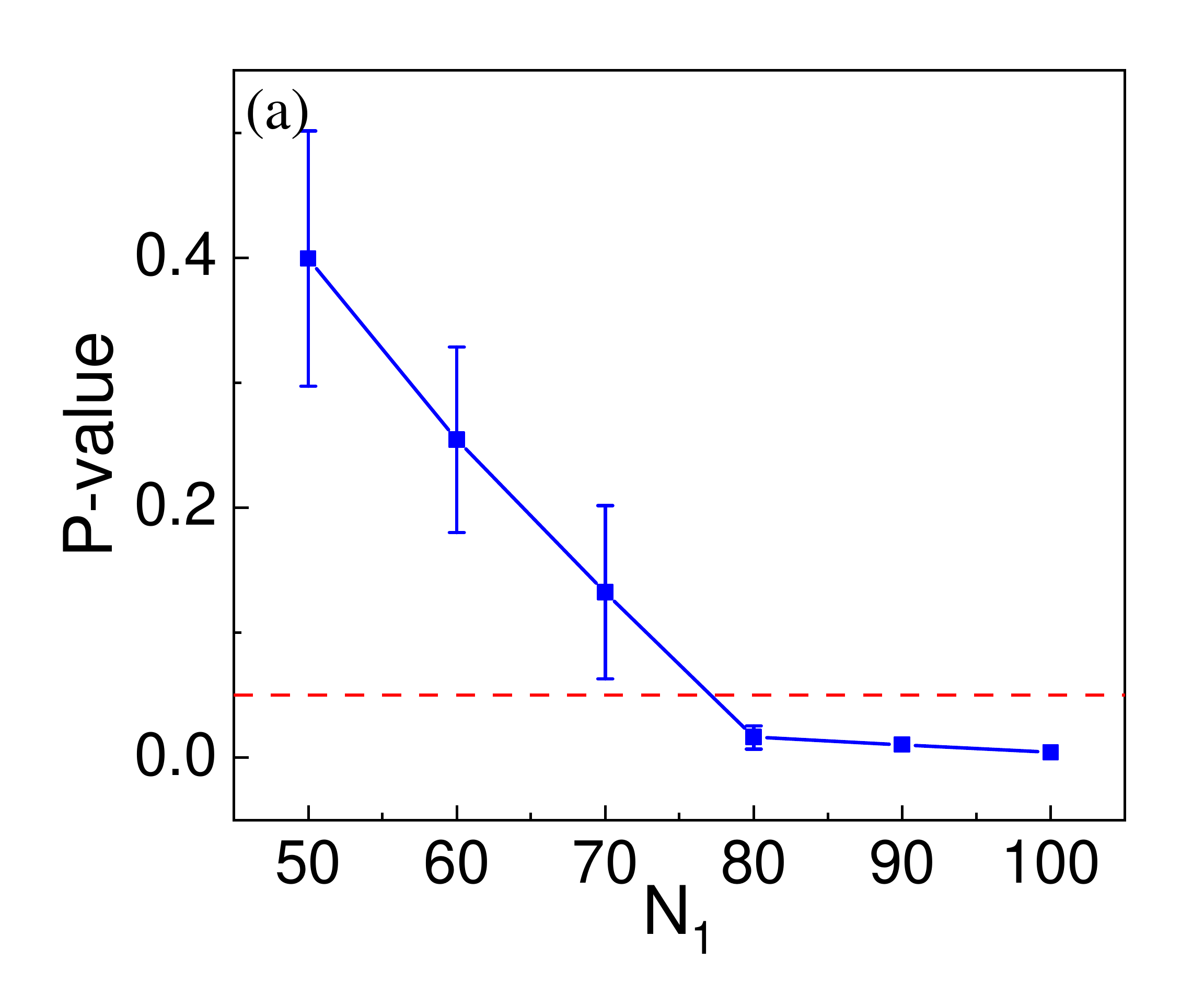}
 \includegraphics[width = 0.45\linewidth]{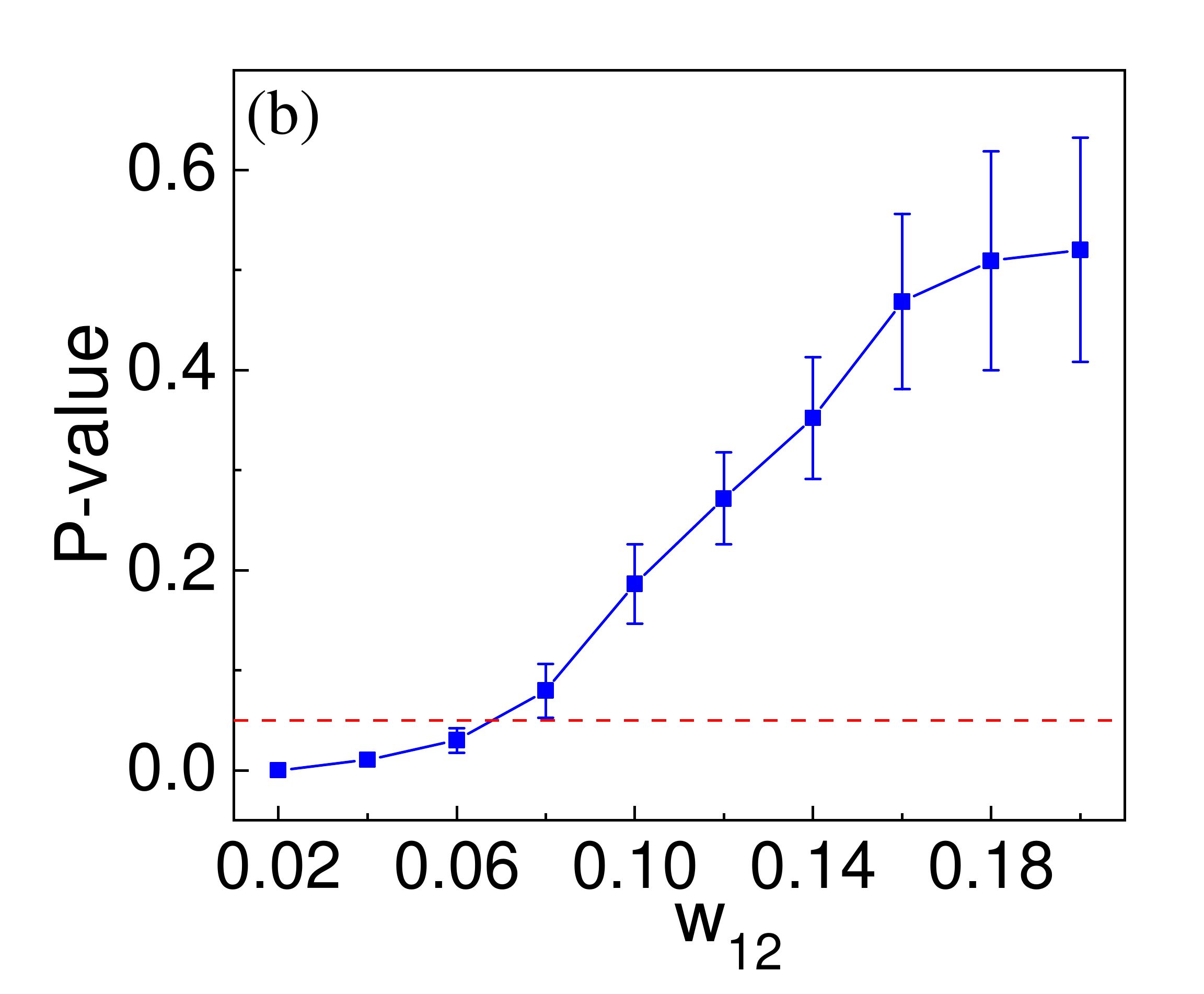}
 \caption{$P$-value as a function of $n_1$ (a) and $w_{12}$ (b) for the unbalanced two-block DCSBM. The dashed lines correspond to the significant level $\alpha=0.05$}
 \label{figdcsbu}
\end{figure}

Next, we apply our algorithm to the DCSBM with unbalanced blocks. We consider the case of two blocks with different size, i.e., $n_1 \neq n_2$. To explore the effect of the community size, we set $w_{12}=w_{21}=0.02$ and $w_{11}=w_{22}=0.2$. As shown in Figure~\ref{figdcsbu}(a), the $P$-value decreases as $n_1$ increases from $50$ to $100$. This is expected since the planted block is easier to detect as $n_1$ grows. In fact, the DCSBM exhibits the block structure for $n_1 \ge 77$. In contrast, we set $n_1=100$ and plot the $P$-values against $w_{12}$ values in Figure~\ref{figdcsbu}(b). One notices the consistent growth of the $P$-value with $w_{12}$. This is also expected since the graph is losing its block structure gradually, especially for $w_{12} \ge 0.068$.

\begin{figure}[t]
\centering
 \includegraphics[width = 0.45\linewidth]{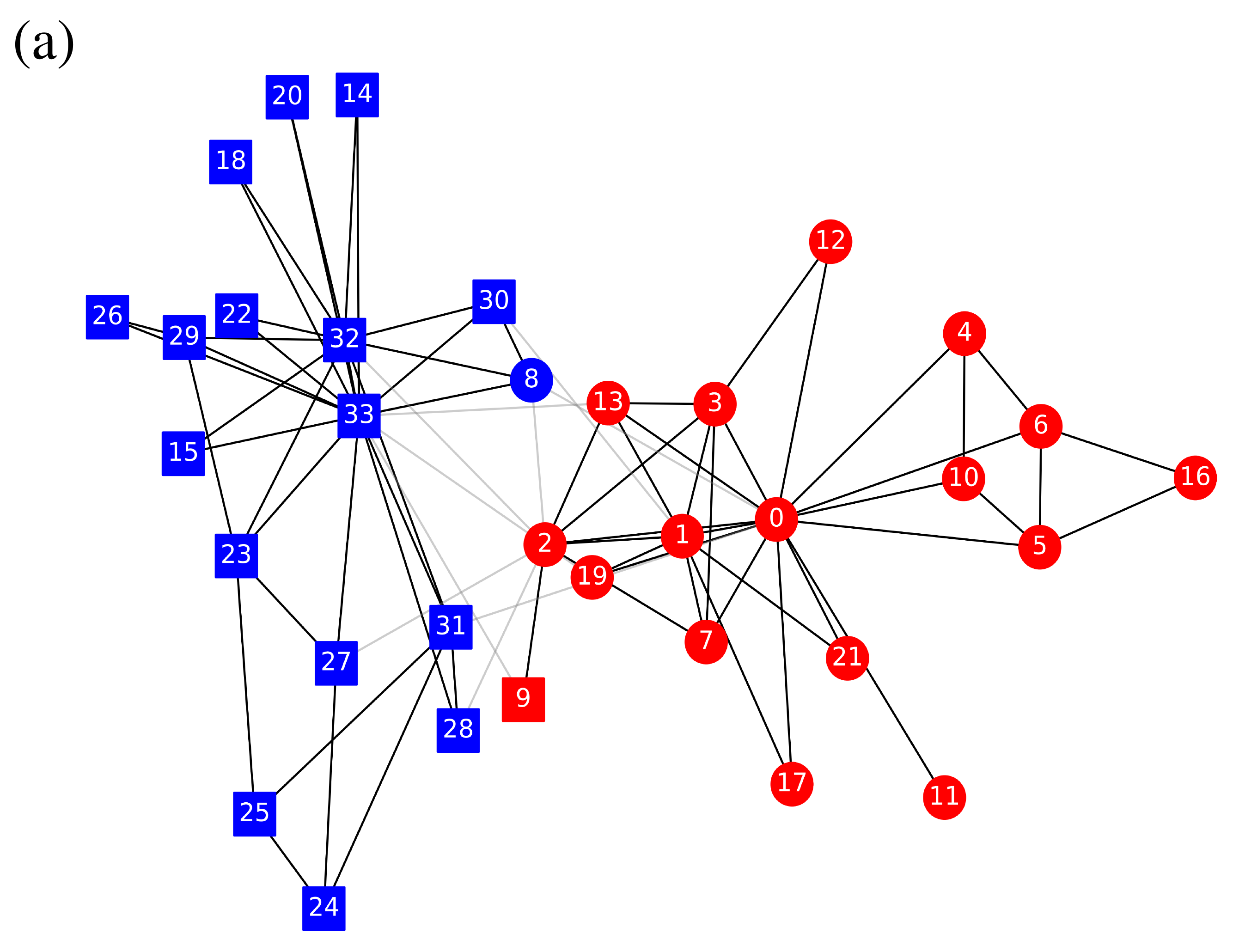}
 \includegraphics[width = 0.45\linewidth]{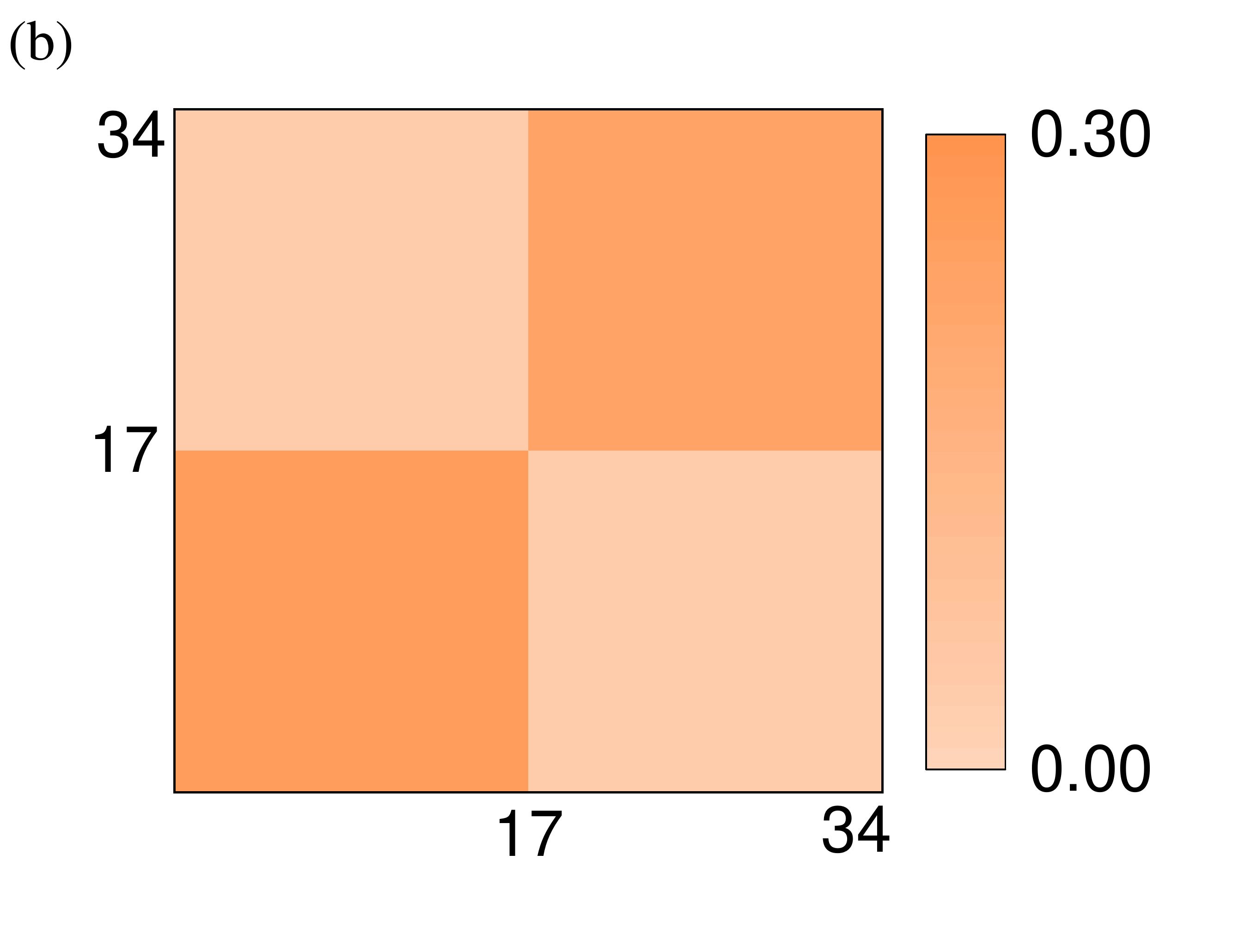}
 \caption{Performance of the hypothesis test algorithm for the karate club: the illustration of the community division (a) and the density plot for the network (b).}
 \label{figkarate}
\end{figure}

\section{Application to empirical networks}

We now turn to applications of our method to empirical networks. The first example, widely studied in literature, is the karate club at an American university recorded over two years by \citet{Zachary77jar}. This is a social network composed of $34$ individuals. Due to a disagreement on class fees between an instructor (node $0$) and an administrator (node $33$), the club split into two different groups and the members of each group are known. Thus, these two groups are considered as the ground truth communities. Applying our algorithm to this network, we find the results shown in Figure~\ref{figkarate}(a). The solid circles and squares represent instructor and administrator clusters, respectively. Except for the misclassification of two vertices (nodes 8 and 9) on the boundary of the two groups, vertices are split in accordance with the known communities. Figure~\ref{figkarate}(b) presents a density image of the adjacency matrix, which also illustrates the block structure.

\begin{figure}[t]
 \centering
 \includegraphics[width = 0.45\linewidth]{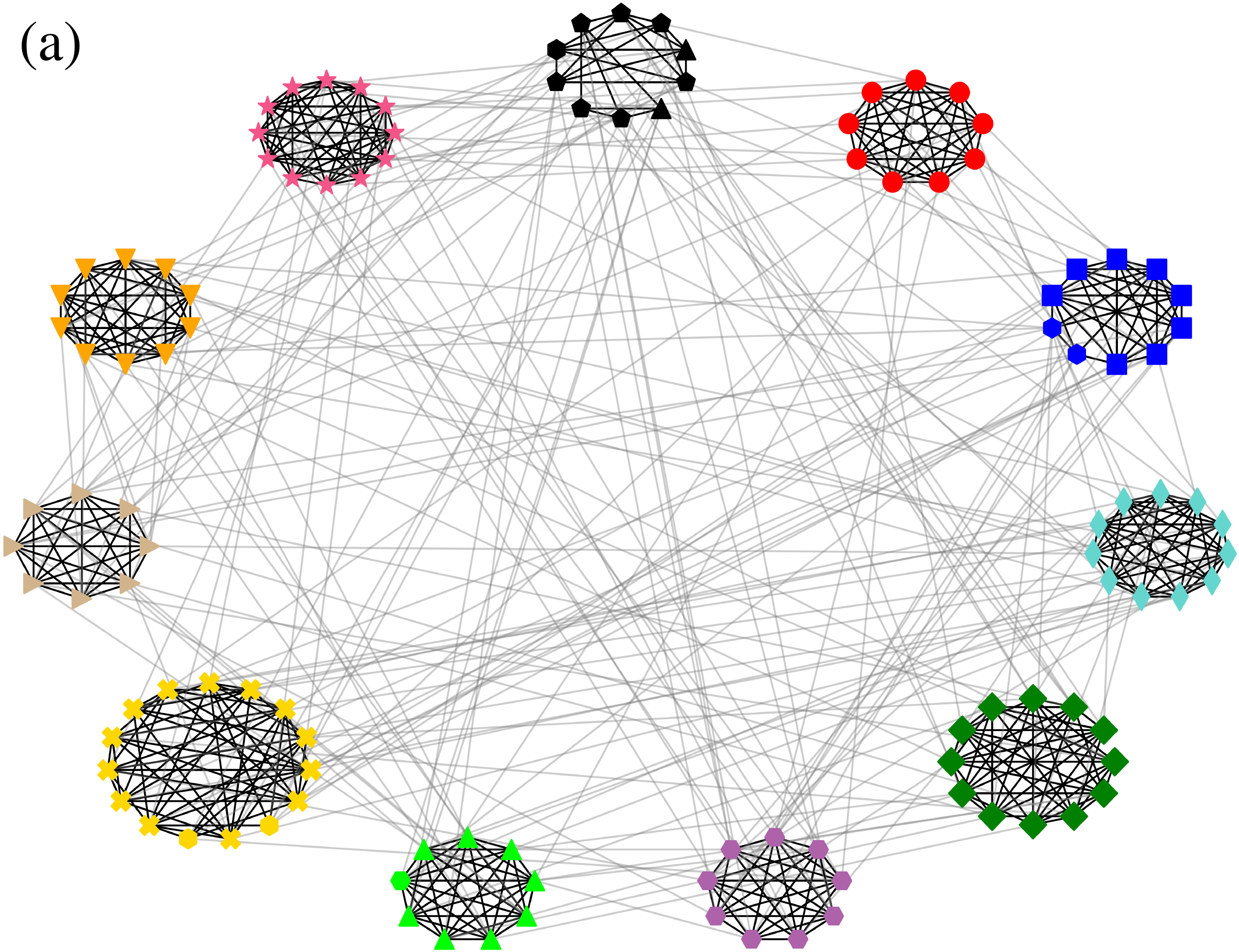}
 \includegraphics[width = 0.45\linewidth]{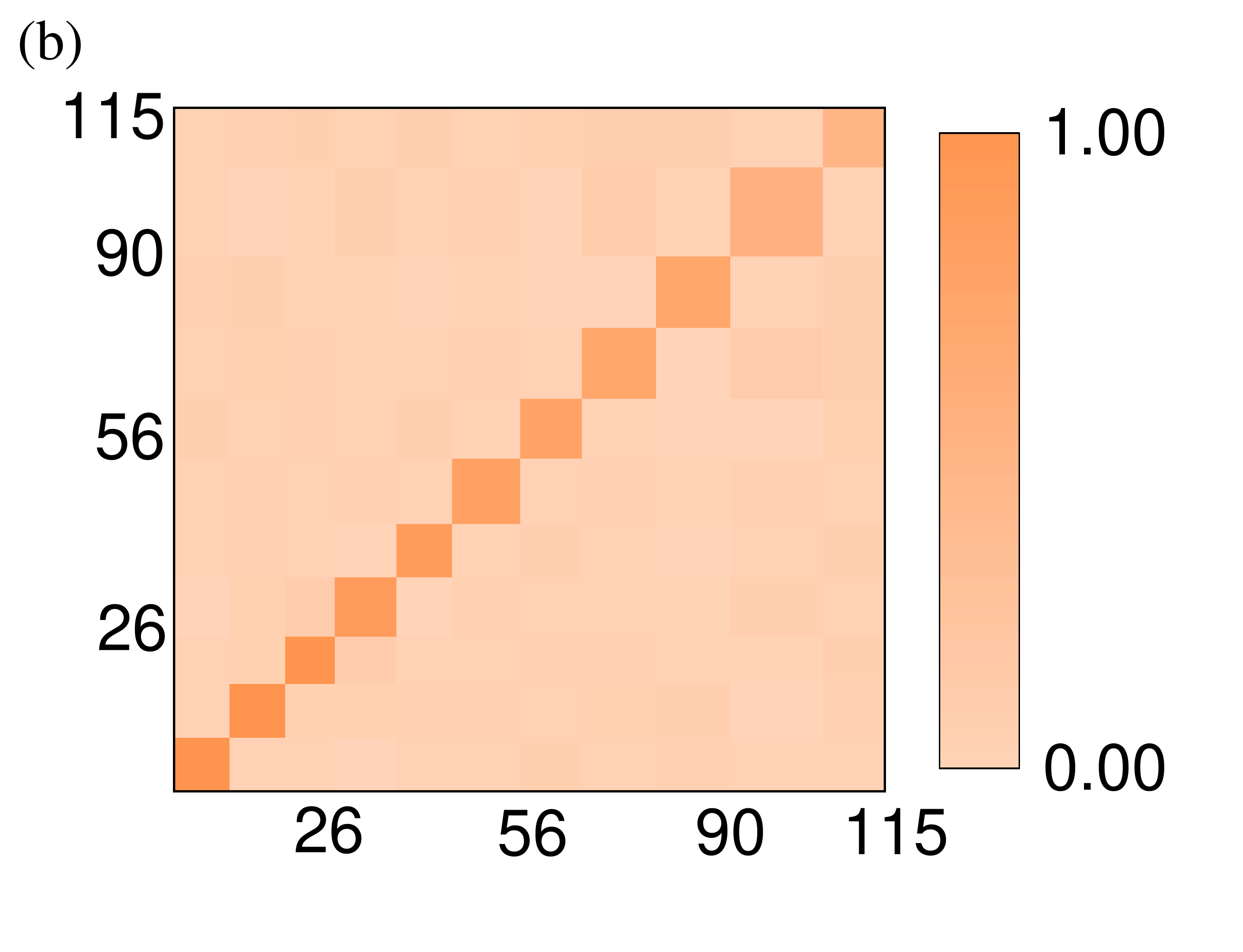}
 \caption{Community division (a) and density matrix (b) for the American college football network.}
 \label{figfoot}
\end{figure}

As a second example in the real world, we consider a network of the American college football network \citep{Girvan02pnas} formed by teams in a league with each vertex representing one team with two teams linked if they have played each other that season. The network consists of $115$ teams in the American College Football Division 1-A in the 2000 season. The teams organized into $12$ conferences and games are more frequent between members of the same conference than between members of different conferences, which leads to a known community structure. Figure~\ref{figfoot}(a) shows the computed community structure by our algorithm. One finds that most teams are correctly grouped with the other teams in their conference except for a few independent teams are settled with conferences they are most closely associated, hence a high degree of agreement. The density plot of the adjacency matrix in Fig.~\ref{figfoot}(b) also elucidate this issue.

To quantitatively compare the results of our algorithm to the ground truth and those of state-of-the-art methods, we introduce two measures: the adjusted Rand index $S_{\mathrm{AR}}$ and $F_1$ score. Given two kinds of classifications $P_a$ and $P_b$, we denote the count of node pairs that classified together in both partitions by $q_{11}$, classified together in $P_a$ but different in $P_b$ by $q_{10}$, different in $P_a$ but classified together in $P_b$ by $q_{01}$, and different in both by $q_{00}$. Noting that $w_{11}+w_{10}+w_{01}+w_{00}=C_n^2=M$, the adjusted Rand index is defined by \citep{Vinh10icml}
\begin{equation}
 S_{\mathrm{AR}}=\frac{w_{11}-\frac{1}{M}\left(w_{11}+w_{10}\right)\left(w_{11}+w_{01}\right)}{\frac{1}{2}\left[\left(w_{11}+w_{10}\right)+\left(w_{11}+w_{01}\right)\right]-\frac{1}{M}\left(w_{11}+w_{10}\right)\left(w_{11}+w_{01}\right)}.
\end{equation}
Another measure comparing $P_a$ and $P_b$ is $F_1$ score, defined as follows \citep{Larsen99kdd}:
\begin{equation}
 F_1=\frac{2\text{precision}(P_a,P_b)\text{recall}(P_a,P_b)}{\text{precision}(P_a,P_b)+\text{recall}(P_a,P_b)}
\end{equation}
with $\text{precision}(P_a,P_b)=|P_a\cap P_b|/|P_b|$ and $\text{recall}(P_a,P_b)=|P_a\cap P_b|/|P_a|$. As shown in Table~\ref{tabcomp}, the number of communities identified for both real networks, $2$ communities in the karate club and $11$ communities in the football network, are much better than those of the state-of-the-art methods. Moreover, the corresponding $S_{\text{AR}}$ and $F_1$ gain highest values, indicating the best alignment with the real communities.

\begin{table*}[h]
\centering
\caption{Comparison of the results of the hypothesis test algorithm to the ground truth and those of the state-of-the-art algorithms.}
\setlength{\tabcolsep}{0.1cm}{
    \begin{tabular}{|c|c|c|c|c|c|c|}
    \hline
    %\hline
    \multicolumn{1}{|c|}
    {}&\multicolumn{3}{c|}{karate club}&\multicolumn{3}{c|}{college football}\cr\cline{1-7}
    &communities &$S_{\text{AR}}$ &$F_1$ &communities &$S_{\text{AR}}$ &$F_1$\cr
    \hline
    hypothesis test              &2 &0.7717 &0.9410 &11 &0.8927 &0.8697 \cr\hline
    Motif-based k-means          &2 &0.6682 &0.9117 &10 &0.7939 &0.8120 \cr\hline
    Clauset~\citep{Clauset04pre} &3 &0.5684 &0.5189 &6  &0.4741 &0.3711 \cr\hline
    Louvain~\citep{Blondel08jsm} &4 &0.4646 &0.3033 &10 &0.8035 &0.6961 \cr\hline
    Infomap~\citep{Rosvall08pnas}&3 &0.5906 &0.5666 &10 &0.8165 &0.6940 \cr\hline
    %\hline
\end{tabular}}
\label{tabcomp}
\end{table*}

\section{Conclusion and future work}

As a mixture model for analyzing structural data, the SBM and its variants have received much interest in detecting communities of networks \citep{Nicola21sm}, among which the DCSBM is particularly well suited for networks with a highly skewed degree distribution.

In this paper, we have proposed a novel hypothesis test for community detection in complex networks. We made two major contributions, the model and the algorithm. In the model aspect, we have defined a graph dissimilarity measure incorporating the vertex distance distribution, the clustering coefficient distribution, and the alpha-centrality distribution. By using this dissimilarity measure between the DCSBM and the DCERG, we put forward a hypothesis testing statistic. In the algorithm aspect, we have devised a two-stage algorithm. We first determined whether the original network is a DCERG. If not we then bipartitioned it until each subgraph is a DCERG. We proposed a new criterion for bipartition incorporating the edge betweenness and the edge clustering coefficient. We applied the algorithm to synthetic and real networks. Overall, the proposed method presents an important advancement over state-of-the-art ones. Therefore, it is feasible to detect communities in networks with broad degree distributions while the actual number of communities is unknown.

There are several avenues for future work. For example, how to measure graph dissimilarity is still an open problem. For networks with higher-order architecture, the new measure beyond pairwise interactions should be taken into account to enhance model capacities \citep{Lacasa21cp}. Furthermore, the Gaussian distribution is a standard choice for the kernel density distribution, but for special interest one may consider other distributions, such as the Epanechnikov distribution widely adopted in financial data analysis. In principle, finding the theoretical distribution for dissimilarity to further reduce the computational complexity is of great importance. Finally, other more sophisticated block models, such as multipartite \citep{Barhen20sm} and dynamic \citep{Bartolucci20prl} models can also benefit from the proposed framework.

%%% Acknowledgements (if any)
%%% ------------------------------------------

\section*{Declaration of Conflicting Interests}
The author(s) declared no potential conflicts of interest with respect to the research,
authorship, and/or publication of this article.

\section*{Funding}
This work was developed within the scope of the project I3N, UIDB/50025/2020 \& UIDP/50025/2020, financed by national funds through the FCT/MEC Portuguese Foundation for Science and Technology. X.-J.X. acknowledges financial support from the Natural Science Foundation of China under Grant No. 12071281.

%%% References (if created by hand).
%%% -----------------------------------------------------------------------------------

\end{document}